\documentclass[12pt]{iopart}
\usepackage{latexsym,amssymb}
\usepackage{graphics}
\usepackage{graphicx}
\usepackage{epsfig}

\renewcommand{\phi}{\ensuremath{\varphi}}

\begin{document}

\title{Construction of the factorized steady state distribution in models of mass
transport}
\author{R.\ K.\ P.\ Zia$^1$, M.\ R.\ Evans$^2$, and Satya\ N.\ Majumdar$^3$}

\address{
$^1$Department of Physics and\\
Center for Stochastic Processes in Science and Engineering,\\
Virginia Tech, Blacksburg, VA 24061-0435, USA\\[0.5ex]
$^2$School of Physics, University of Edinburgh,\\
Mayfield Road, Edinburgh EH9 3JZ, UK\\[0.5ex]
$^3$ Laboratoire de Physique Theorique et Modeles
Statistiques,\\ Universite Paris-Sud, Bat 100, 91405, Orsay-Cedex,
France.}

\date{Jul 28th , 2004}

\begin{abstract}
For a class of one-dimensional mass transport models we present a
simple and direct test on the chipping functions, which define the
probabilities for mass to be transferred to neighbouring sites, to
determine if the stationary distribution is factorized. In cases where the
answer is affirmative, we provide an explicit method for constructing
the single-site weight function. As an illustration of the power of
this approach, previously known results on the Zero-range process and
Asymmetric random average process are recovered in a few lines. We
also construct new models, namely a generalized Zero-range process and
a binomial chipping model, which have factorized steady states.
\end{abstract}


\maketitle

In a recent publication \cite{EMZ}, we investigated a class of mass
transport models on a ring (one-dimensional lattice with periodic
boundary condition).  The class encompasses both discrete and
continuous masses and discrete and continuous time dynamical rules for
the transfer of mass between neighbouring sites.  The Zero-range
process (ZRP) \cite{Spitzer,MRE00} and Asymmetric random average process
(ARAP) \cite{KG,RM}
correspond to special cases of this class.  We derived a necessary and
sufficient condition for the steady state to factorize which dictates
an appealingly simple, yet general, form for the chipping functions
($\phi(\mu|m)$ in equation \ref{answer!} below) which
define the probabilities with which mass is transferred from site to
site. Though the form of this condition might be elegant, it is an
``implicit test'' for the chipping functions.  

In the present work we
formulate a simple explicit test for the chipping functions, as
well as a straightforward method for constructing the single-site 
weight function
(products of which form the steady state distribution). In this sense,
we provide a complete solution to this class of mass transport
models. This note may be considered as a sequel to \cite{EMZ}, where
the reader will find the motivation for the model and 
details of the previous analysis.

Following the notation of \cite{EMZ}, our system consists of masses $m_i$ at
site $i=1\ldots L$ on a ring. At each time step, a mass $\mu _i$ (drawn from
a distribution $\phi (\mu _i|m_i)$) is `chipped off' $m_i$ and moved to site 
$i+1$. After long times, the system settles into a time-independent state,
with distribution (unnormalized probabilities) $F(m_1,...,m_L)$. The central
result of \cite{EMZ} (equation 15) is a necessary and sufficient condition
on $\phi (\mu |m)$ for $F$ to factorize: $F=f(m_1)...f(m_L)$, namely, if and
only if $\phi (\mu |m)$ can be expressed in the form {\ 
\begin{equation}
\phi (\mu |m)=\frac{v\left( \mu \right) w\left( m-\mu \right) }{\left[
v*w\right] \left( m\right) } \,\, ,  \label{answer!}
\end{equation}
where }$v$ and $w$ are two non-negative functions. Further, $f(m)$, the single
site weight is given by the convolution in the denominator, i.e., 
\begin{eqnarray}
f\left( m\right) =\left[ v*w\right] \left( m\right) \equiv \int_0^m %
\ensuremath{\mathrm{d}}\mu\, v\left( \mu \right) w\left( m-\mu \right) \,\,.
\label{fm}
\end{eqnarray}

Typically, a mass transport model is motivated by a specific chipping function
 $%
\phi $. As a result, it is not particularly easy to see if it satisfies
condition ({\ref{answer!}}), the form of which, though elegant, is
``implicit'' . We now take the next simple step and turn this into an
``explicit'' test for $\phi $. As in \cite{EMZ}, it is clearer to regard the
two variables in $\phi $ as $\mu $ and $\sigma \equiv m-\mu $. 
Note that the condition for factorization in Eq. (\ref{answer!}) is equivalent
to asking if
\begin{equation}
\left. \frac \partial {\partial \mu }\right| _\sigma \left. \frac \partial
{\partial \sigma }\right| _\mu \ln \phi \left( \mu |\mu +\sigma \right)
\label{test}
\end{equation}
is a function of $\mu + \sigma$ alone
(i.e., $m$, and no other dependence $\mu $, say). 
Thus, if we have factorization, the quantity in Eq. (\ref{test})
is a function of $m$ alone and let us call it  $h\left( m\right)$. 
That this is a necessary and sufficient condition equivalent to
(\ref{answer!}) can then be seen by integrating (\ref{test})
with respect to $\mu$ and $\sigma$, which yields $\phi$ of the form
(\ref{answer!}).
Moreover, integrating twice
with respect to $m$ and then exponentiating, we have explicitly 
\begin{eqnarray}
f\left( m\right) =\exp \left[ -\int^m \ensuremath{\mathrm{d}} m^{\prime
}\int^{m^{\prime }}\ensuremath{\mathrm{d}} m^{\prime \prime }h\left(
m^{\prime \prime }\right) \right] \,\,.  \label{fm-exp}
\end{eqnarray}
Note that there are two arbitrary integration constants in this
construction, leading to an overall amplitude and exponential factor $a^m$.
These are precisely the two ``degrees of freedom'' we encountered while
defining $f(m)$ in \cite{EMZ}.

This formalism thus 
provides a simple test (\ref{test}) to see if any specific
mass transport model admits a factorized steady state and, if so, a recipe for the 
associated single-site weight (\ref{fm-exp}). As an example of the applicability of ({%
\ref{test}) and (\ref{fm-exp}}), we consider the case of the ARAP \cite
{KG,RM,CLMNW,ZS1} which is a class of mass transport models defined by a chipping
function of the form $\phi (\mu |m)=\psi(\mu/m)/m$ i.e. a random fraction $%
r=\mu/m$ chips off at each update. In this case (\ref{test}) gives 
\begin{equation}
- \frac{r(1-r)}{m^2}\frac{\ensuremath{\mathrm{d}} K(r)}{\ensuremath{%
\mathrm{d}} r} - \frac{1-2r}{m^2}K(r)+ \frac{1}{m^2}  \label{hARAP}
\end{equation}
where 
\begin{equation}
K(r) = \frac{1}{\psi(r)}\frac{\ensuremath{\mathrm{d}} \psi(r)}{%
\ensuremath{\mathrm{d}} r}\;.
\end{equation}
Thus, the condition that (\ref{test}) depends on $m$ alone implies 
$h(m) \propto 1/m^2$ and
\begin{equation}
\frac{\ensuremath{\mathrm{d}} }{\ensuremath{\mathrm{d}} r}
\left[r(1-r)K(r)\right]= \mbox{constant}
\end{equation}
the solution of which is 
\begin{equation}
\psi(r) = C r^p(1-r)^q\;,
\end{equation}
where the constant $C$ is fixed by the normalisation condition 
\begin{equation}
\int_0^1 \ensuremath{\mathrm{d}} r \, \psi(r) =1\;.
\end{equation}
This condition yields 
\begin{equation}
\psi(r) = \frac{r^p(1-r)^q}{B(p+1,q+1)}  \label{psiARAP}
\end{equation}
where $B(p+1,q+1)$ is the usual Beta function and $p$,$q > -1$. The function 
$h(m)$ given by (\ref{test}) is from (\ref{hARAP}) 
\begin{equation}
h(m)= \frac{1}{m^2}\left\{ -\frac{\ensuremath{\mathrm{d}}}{%
\ensuremath{\mathrm{d}} r}\left[ r(1-r) K(r)\right] +1\right\} = \frac{1+p+q%
}{m^2}
\end{equation}
and, integrating twice (\ref{fm-exp}), we retrieve the result: $f(m)\propto m^{1+p+q}\,\,$.
Thus (\ref{psiARAP}) is the most general chipping function of $r$ that gives
rise to a factorized steady state. This proves in a direct way a result of 
\cite{ZS2}.

For models with discrete masses such as the ZRP, the derivatives above
become differences. To be careful, let us write the chipping rates as 
\[
\phi (\mu |m)=\sum_{n=1}^\infty \sum_{\ell =0}^n\varphi _{\ell ,n}\delta
(\mu -\ell )\delta (m-n)\,\,.
\]
The factorization condition for continuous masses in Eq. (\ref{answer!})
has an equivalent discrete analogue,
\begin{equation}
\phi_{l,n} = \frac{v_l w_{n-l}}{f_n},
\label{discrete1}
\end{equation}
where $f_n$'s are single-site weights. 
The factorization test can now be rephrased in terms of the cross ratio
\begin{equation}
R\left( \ell ,n\right) \,\equiv \frac{\varphi _{\ell +1,n+2}\varphi _{\ell
,n}}{\varphi _{\ell +1,n+1}\varphi _{\ell ,n+1}}\,\,\,,  \label{R}
\end{equation}
defined when \emph{all} of the $\varphi $'s are positive. 
If the factorization condition in Eq. (\ref{discrete1}) holds 
then the cross ratio $R$ is clearly independent of $l$ and depends only on $n$.
Thus, a
factorized steady state exists \emph{if and only if} $R(l,n)$ is independent of $%
\ell $. In this case $R(n)$ will be given in terms of the single-site
weights $f_n$ in 
\begin{equation}
f(m)=\sum_{n=1}^\infty f_n\delta (m-n)  \label{zrpss}
\end{equation}
as 
\begin{equation}
R(n)=\frac{f_{n+1}^2}{f_n\,f_{n+2}}\;,
\end{equation}
which yields the recursion 
\begin{eqnarray}
\frac{f_{n+2}}{f_{n+1}}=\frac{1}{R(n)}\frac{f_{n+1}}{f_n}\;.  \label{recur1}
\end{eqnarray}
Iterating (\ref{recur1}) yields 
\begin{equation}
\frac{f_{n+2}}{f_{n+1}}=\left[ \prod_{m=0}^n\frac{1}{R(m)}\right] 
\frac{f_1}{f_0}\;,
\end{equation}
which one can iterate again to obtain 
\begin{equation}
f_n=\left( f_0 \right) \left( \frac{f_1}{f_0}\right) ^n
\left[ \prod_{m=0}^{n-2}\frac{1}{R(m)^{n-m-1}} \right] 
\quad \mbox{for}\quad n\geq
2\;.  \label{fn}
\end{equation}
Of course, this is the discrete version of ``integrating twice''. The two
arbitrary constants (overall amplitude $f_0$ and exponential amplitude $%
a^n=(f_1/f_0)^n$) are again explicitly displayed.

As an illustration consider the ZRP where only unit masses can chip off with
probability $u(n)$ where $n$ is the mass at the site \cite{MRE00}. In this
case, we see that there is \emph{only one} cross ratio, namely for $l=0$, 
\begin{equation}
R\left( 0,n\right) =\frac{u\left( n+2\right) \left[ 1-u\left( n\right)
\right] }{u\left( n+1\right) \left[ 1-u\left( n+1\right) \right] }\,\,.
\label{zprR}
\end{equation}
Since this $R$ is automatically ``independent of $\ell $,'' we
immediately recover the conclusion: the ZRP admits a
factorized steady state. Further, it is straightforward to retrieve
the $f_n$'s from ({\ref{fn}) as
\begin{equation}
f_n=\left( \frac{f_0}{1-u(n)}\right) \left( \frac{f_1u(1)}{f_0}\right)
^n\left[ \prod_{m=1}^n\frac{1-u(m)}{u(m)}\right] \,\,,
\end{equation}
which recovers the result of \cite{EMZ} originally derived by a more
complicated approach in \cite{MRE97}. Note that we have again displayed the
two arbitrary constants with }$\left( ...\right) $ brackets.

As a simple example of
a new model with  a factorized steady state we define the
binomial chipping model, in which the
mass at each site is discrete, $m=0,1,2,\ldots $. The chipping is
specified by the following kernel,
\begin{equation}
\phi _{\ell ,n} = {n\choose \ell }p^\ell (1-p)^{n-\ell },  \label{binom1}
\end{equation}
where $0\le p\le 1$ is a parameter and $\ell =0,1,\dots n$. One can
understand the model by interpreting $n$ as the number of unit masses
at a site, each of which move independently with probability $p$ at
each time-step. One finds
\begin{equation}
R(\ell ,n)=\frac{n+2}{n+1}\;
\end{equation}
for \emph{all }$\ell $. Since manifestly this is
independent of $\ell $  the steady state factorizes and (\ref{fn})
yields 
\begin{equation}
f_n=f_0\left( \frac{f_1}{f_0}\right) ^n\frac 1{n!}\;.
\end{equation}
Note that, apart from the ``irrelevant'' factors, the single site weight
here is extremely simple: $1/n!$.

As a more involved example of constructing a new model with a
factorized steady state, we consider a generalized zero-range process
where mass chunks of size one or two can chip off at each time step
with probabilities $u_1(n)$ and $u_2(n)$ respectively. In this case we
have two cross ratios
\begin{eqnarray}
R\left( 0,n\right)  &=&\frac{u_1\left( n+2\right) \left[ 1-u_1\left(
n\right) -u_2\left( n\right) \right] }{u_1\left( n+1\right) \left[
1-u_1\left( n+1\right) -u_2\left( n+1\right) \right] }
\quad\mbox{for}\quad n\geq 0 \label{R0}\\
R\left( 1,n\right)  &=&\frac{u_2\left( n+2\right) u_1\left( n\right) }{%
u_2\left( n+1\right) u_1\left( n+1\right) }
\quad\mbox{for}\quad n\geq 1 \;.
\label{R1}
\end{eqnarray}
If we demand that the stationary state factorizes, then we must
have $R\left( 0,n\right) =R\left( 1,n\right) $
which reduces to
\begin{equation}
\frac{u_2(n+1)(1-u_1(n)-u_2(n))}{u_1(n+1) u_1(n)} = A
\quad\mbox{for}\quad n\geq 1
\label{Adef}
\end{equation}
where $A$ is a positive constant independent of $n$.
In terms of the ratio 
\begin{equation}
\rho _n\equiv \frac{u_2(n)}{u_1(n)}\,\,,  \label{yn}
\end{equation}
(\ref{Adef}) becomes
\begin{equation}
\rho _{n+1}\left( \frac{1-u_1(n)}{u_1(n)}-\rho _n\right) =A \;. \label{nlrecur}
\end{equation}
 Though this condition is in the form of a 
\emph{nonlinear} recursion, we can linearize it by changing variables to $y_n
$ via 
\begin{equation}
\rho _n=A\frac{y_{n-1}}{y_n}  \label{cond} \;.
\end{equation}
 Then (\ref{nlrecur}) becomes a linear second-order
recursion for the $y$'s:
\begin{equation}
y_{n+1}=\frac{1-u_1(n)}{u_1(n)} y_{n}-A y_{n-1}\;,  \label{wm}
\end{equation}
with initial conditions $y_0 =0$ and $y_1=1$.
Thus we can find $y_n$ in terms of arbitrary $u_1(n)$ and $A$. 
At the same time,
these $y$'s also fix the ``allowed'' chipping rates for two mass units, $%
u_2(n)$, in terms of $u_1(m)$, $m\leq n$ and one free parameter $A$,
through the relations (\ref{yn}) and (\ref{cond}):
\begin{equation}
u_2(n) = Au_1(n) \frac{y_{n-1}}{y_n}\;.
\end{equation}

Of course, the
single-site weights can also be found using (\ref{fn}) 
and (\ref{R0})
\begin{equation}
f_n=\left( \frac{f_0}{1-u_1(n)-u_2(n)}\right) \left( \frac{f_1 u_1(1)}{f_0}%
\right) ^n\left[ \prod_{m=1}^n \frac{1-u_1(m)-u_2(m)}{u_1(m)}\right] 
\end{equation}
which may also be written in terms of $y_n$ as
\begin{equation}
f_n = f_0 \left( \frac{f_1 u_1(1)}{f_0}%
\right)^n \frac{y_n}{u_1(n)}\;.
\end{equation}

Finally, we remark that the constructive method presented here easily
adapts to the case of continuous time 
(or random sequential) dynamics specified
by the rates per unit time, $\gamma(\mu|m)$, at which mass $\mu$ chips
off mass $m$. It is shown in \cite{EMZ} that the necessary and
sufficient condition for a factorized steady state becomes
\begin{equation}
\gamma(\mu|m) = \frac{x(\mu) w(m-\mu)}{w(m)}
\end{equation}
in which case the single-site weights become $f(m)= w(m)$.  Thus the
test for and construction of the factorized steady states is identical
to that of the discrete time case with $\phi(\mu|m)$ replaced by
$\gamma(\mu|m)$.

\ack

This research is supported in part by the US National Science Foundation
through DMR-0088451 and DMR-0414122. 

\end{document}